\documentclass[aps,prb,reprint,superscriptaddress]{revtex4-2}
\usepackage{color}
\usepackage{bm}
\usepackage{braket}
\usepackage{ulem}
\usepackage{graphicx}
\usepackage{hyperref}

\hypersetup{
	colorlinks=true,
	anchorcolor=blue,
	linkcolor=blue,
    citecolor=blue,
    urlcolor=blue
}

\begin{document}

\title{Spin Transport in Normal Metal-Ising Superconductor Junction}

\author{Yi-Xin Dai}
\affiliation{International Center for Quantum Materials, School of Physics, Peking University, Beijing 100871, China}
\affiliation{Beijing Academy of Quantum Information Sciences, West Bld.\#3,No.10 Xibeiwang East Rd., Haidian District, Beijing 100193, China}
\author{Yue Mao}
\affiliation{International Center for Quantum Materials, School of Physics, Peking University, Beijing 100871, China}
\author{Qing-Feng Sun}
\email{sunqf@pku.edu.cn}
\affiliation{International Center for Quantum Materials, School of Physics, Peking University, Beijing 100871, China}
\affiliation{Beijing Academy of Quantum Information Sciences, West Bld.\#3,No.10 Xibeiwang East Rd., Haidian District, Beijing 100193, China}

\date{\today}

\begin{abstract}
The combination of spin-orbit coupling and superconductivity induces unconventional spin-triplet correlation in Ising superconductors. 
We theoretically investigate the spin transport through a normal metal-Ising superconductor junction, showing that Ising superconductors also have the characteristic of spin superconductivity.
Due to the existence of spin-triplet Cooper pairs, not only charge supercurrent but also spin supercurrent can transport in Ising superconductors.
We analyze the transport process in the junction which is mainly contributed by the equal-spin Andreev reflection and spin-flip reflection, and calculate the spin conductance and the spin injection efficiency under different conditions.
Our findings broaden the boundary of spin superconductivity and reveal the potential applications of Ising superconductors in spintronics, especially in controlled long-distance dissipationless spin transport.
\end{abstract}

\maketitle

\section{\label{intro}Introduction}
The manipulating of spin transport is one of the most significant topics
in spintronics \cite{zutic_Spintronics_2004,eschrig_Spinpolarized_2015}.
Due to the vectorial feature of spin operators, in general,
the spin current is a tensor and can be driven by a
spin bias \cite{wang_Spinbattery_2004,sun_Persistent_2008}.
Such spin bias, especially in metals, splits the chemical potentials
of electrons with opposite spin.
Therefore, electrons with opposite spin move oppositely,
generating a nonzero spin current.
However, this transport suffers from the suppression of spin relaxation and dephasing mechanisms, causing the decreasing of the spin lifetime and transport distance, which prevents it from wide applications \cite{zutic_Spintronics_2004,avsar_Colloquium_2020,sierra_Van_2021}.
To this end, the newly proposed state of spin supercondcutivity \cite{sun_Spin_2011,liu_Spontaneous_2012,sun_Spinpolarized_2013,bao_Ginzburg_2013,lv_GinzburgLandautype_2017,jiang_SpinTriplet_2020,wu_NonAbelian_2022} can effectively overcome such difficulty.
As a counterpart of the charge superconductivity, the spin superconductivity was proposed in the research of ferromagnetic graphene.
The condensation of spin-triplet electron-hole pairs can form a superfluid state to achieve zero spin resistance while the whole system remains a charge insulator \cite{sun_Spin_2011}.
As an analogy to the charge superconductor, spin superconductor can mediate dissipationless transport of spin current and has an electric Meissner effect as well as spin-current Josephson effect \cite{sun_Spin_2011,nakata_Josephson_2014,bozhko_Supercurrent_2016,zhang_Dissipationless_2022}, both of which can be described by the Ginzburg-Landau-type equations \cite{sun_Spin_2011,bao_Ginzburg_2013,lv_GinzburgLandautype_2017,mao_Spin_2022}.
Since its appearance, the domain of spin superconductivity has been broadening, including spin-triplet exciton condensation, spin superfluidity in magnetic insulators, and so on \cite{takei_Superfluid_2014,skarsvag_Spin_2015,takei_Nonlocal_2015,takei_Spin_2016,flebus_TwoFluid_2016}.

Since the discovery of graphene, the emerging two-dimensional materials have offered a unique platform for spintronics, providing opportunities for future applications \cite{tombros_Electronic_2007,sierra_Van_2021}.
Among them, two-dimensional transition metal dichalcogenides (TMDs) have gained growing interests due to their strong spin-orbit coupling (SOC) in both conduction and valence bands \cite{manzeli_2D_2017}.
In recent years, experiments have observed exotic superconducting state in two-dimensional TMDs films, which exhibits an enhancement of in-plane upper critical field far beyond the Pauli paramagnetic limit \cite{lu_Evidence_2015, saito_Superconductivity_2016, xi_Ising_2016,delabarrera_Tuning_2018}.
Both theories and experiments have revealed that this superconductivity is due to the inter-valley pairing protected by the SOC-induced Zeeman-type spin-valley locking, namely Ising supercondcutivity \cite{xiao_Coupled_2012,rosner_Phase_2014,lu_Evidence_2015,saito_Superconductivity_2016,xi_Ising_2016}.
Further researches found that there exist spin-triplet correlations in Ising superconductors \cite{zhou_Ising_2016,mockli_Magneticfield_2019,haim_Signatures_2020}.
Such spin-triplet correlations can affect the transport process in Ising superconductors, ensuring the appearance of equal-spin Andreev reflection.
Unlike normal Andreev reflection, in the equal-spin Andreev reflection, the incident electrons and reflected holes are in the same spin sub-band, injecting spin-triplet Cooper pairs into the Ising superconductor.
It has been found that in ferromagnet-Ising superconductor junctions, there exists equal-spin Andreev reflection which has a magnetoanisotropic period $\pi$ different from $2\pi$ in the conventional magnetoanisotropic system \cite{lv_Magnetoanisotropic_2018}.
Similarly, in Ising superconductor Josephson junctions with a ferromagnetic center, the equal-spin Andreev reflection also results in a $\pi$-period magnetoanisotropic behavior in the current-phase difference relations and induces a switch effect and $0-\pi$ transition for the charge transport \cite{cheng_Switch_2019,tang_Controlling_2021}.
Besides, equal-spin cross Andreev reflection is also found in junctions where the Ising superconductor is sandwiched by two ferromagnets \cite{lu_Equalspin_2022}.
As the spin-triplet Cooper pairs possess both charge $2e$ and spin $\hbar$, the unconventional Ising superconductor should not only be the charge superconductor but also be viewed as a superconductor of spin \cite{mao_Spin_2022}.
Therefore, with the help of the equal-spin Andreev reflection, the spin current would be able to flow dissipationlessly through the Ising superconductor.

In this paper, we theoretically investigate the spin transport through a normal metal-Ising superconductor junction, showing that Ising superconductors are also spin superconductors.
Under a spin bias \cite{wang_Spinbattery_2004}, a spin current can flow through the junction.
Two spin transport processes, the equal-spin Andreev reflection and spin-flip reflection, occur at the interface of the junction.
They correspond to injecting the spin into Ising superconductor and dissipating spin at the interface, respectively.
Using nonequilibrium Green's function method, we obtain the corresponding coefficients.
Then we investigate the spin current of the system and the spin injection efficiency under various parameters.
These results show the possibility to detect spin superconductivity in Ising superconductors and its potential applications for controlled dissipationless long-distance spin transport.

The rest of the paper is as follows: In Sec. \ref{model}, we give the Hamiltonian of the normal metal-Ising superconductor junction and derive the formula for the spin transport process.
In Sec. \ref{equal-spin} and \ref{spin-flip}, we study the equal-spin Andreev reflection and spin-flip reflection, respectively.
They are the main processes contributing to the spin transport.
In Sec. \ref{spin-cond}, we further consider the impact of spin bias and calculate the total spin conductance of the system.
Finally, a brief summary is presented in Sec. \ref{diss}.

\section{\label{model}Model and formula}
\begin{figure}
    \includegraphics[width=0.9\linewidth]{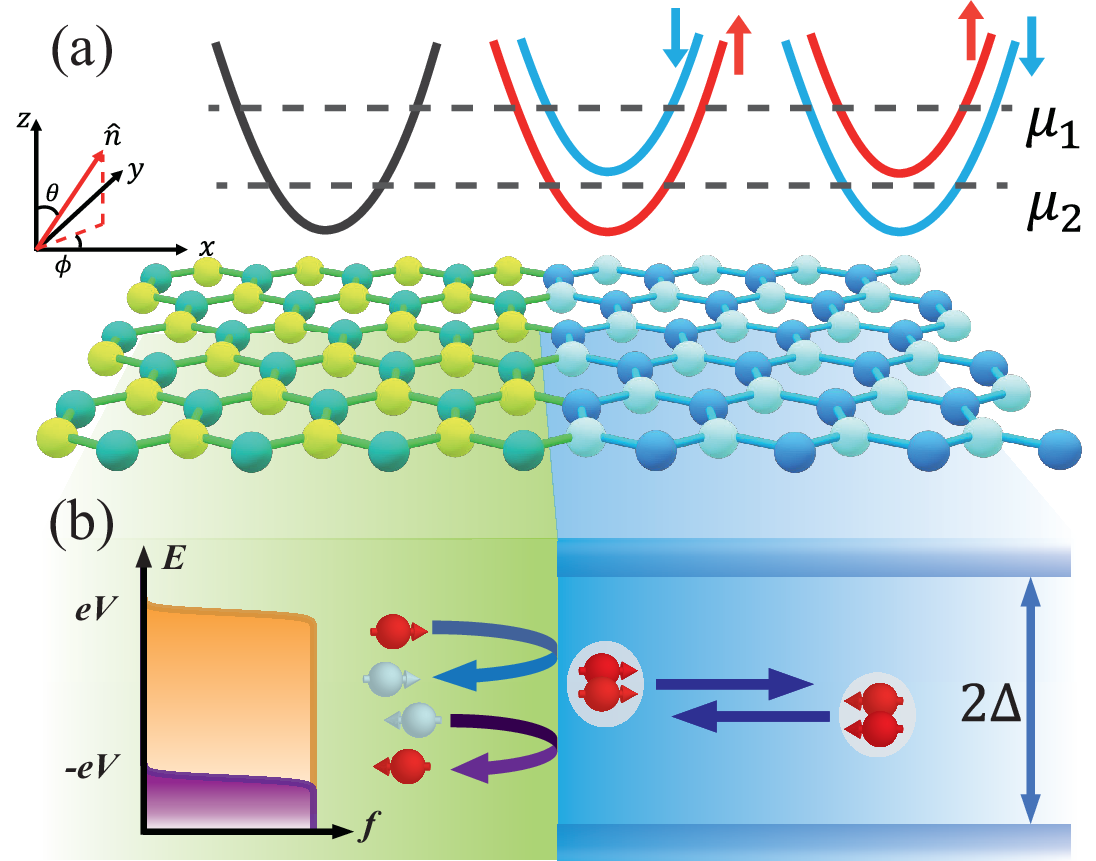}
    \caption{\label{p_model}
    (a) Schematics of the normal metal-Ising superconductor junction with the normal state energy band of the heterostructure in the upper part. 
    While the left lead remains a metal state, for different Fermi energy, the right lead would be in different phases.
    Here $\mu_1$ and $\mu_2$ indicate the Fermi energy for double-band and single-band cases, respectively.
    Besides, $\hat{\bm n}$ denoting the direction of spin bias.
    (b) The spin transport process under a spin bias.
    The spin bias drives a spin supercurrent injected into the Ising superconductor by the equal-spin Andreev reflection, as spin-triplet Cooper pairs carrying opposite spin flow oppositely in the Ising superconductor.
    }
\end{figure}

We consider the system depicted in Fig. \ref{p_model}(a), consisting of a normal metal coupled to an Ising superconductor.
The system is described by the following tight-binding Hamiltonian $H_{tot}=H_N + H_{IS} + H_C \nonumber$, with \cite{lv_Magnetoanisotropic_2018}
\begin{eqnarray}
    &H_N& = \sum_{\bm{m}s} (
            \epsilon-E_L + \lambda_{\bm{m}}\beta
        ) a^\dagger_{\bm{m}s} a_{\bm{m}s}
        + \sum_{\langle\bm{mn}\rangle s}
            t a^\dagger_{\bm{m}s} a_{\bm{n}s} ,\nonumber\\
    &H_{IS}& = \sum_{\bm{m}s} (
            \epsilon-E_R + \lambda_{\bm{m}}\beta
        ) b^\dagger_{\bm{m}s} b_{\bm{m}s}
        + \sum_{\langle\bm{mn}\rangle s}
            t b^\dagger_{\bm{m}s} b_{\bm{n}s} \nonumber\\
    &+& \sum_{\langle\langle\bm{mn}\rangle\rangle ss'}
        i\beta_s \nu_{\bm{mn}} \sigma^z_{ss'}
        b^\dagger_{\bm{m}s} b_{\bm{n}s'}
        + \sum_{\bm{m}} (
            \Delta b^\dagger_{\bm{m}\uparrow} b^\dagger_{\bm{m}\downarrow}
            + h.c.
        ),\nonumber\\
    &H_C& = \sum_{\bm{mn}s} t_c(\bm{m},\bm{n}) a^\dagger_{\bm{m}s} b_{\bm{n}s} + h.c. ,
\end{eqnarray}
where $H_N, H_{IS}, H_C$ represent the Hamiltonians of normal metal, Ising superconductor, and their coupling, respectively.
$a_{\bm{m}s}$ and $b_{\bm{n}s}$ are electron annihilation operators in the normal metal and Ising superconductor, with $\bm{m,n}$ denoting the discrete sites and $s=\uparrow, \downarrow$ denoting the electron spins.
Here both the normal metal and Ising superconductor are hexagonal lattice, with $\epsilon-E_{L/R}$ denoting the on-site energy and $\lambda_{\bm{m}}\beta=\pm \beta$ for the energy difference between A-B sublattice.
Moreover, the on-site energy $E_{L/R}$ can be modulated by the gate voltage in experiments \cite{shi_Superconductivity_2015,saito_Superconductivity_2016}
so that the right lead can be regulated to either single-band or double-band Ising superconducting state with the left lead remaining the metal state 
[see the energy band in Fig. \ref{p_model}(a)].
The SOC in Ising superconductor is described by the second-nearest neighbor hopping term with $\nu_{\bm{mn}}=\pm 1$ depending on the orientation of the two sites ${\bm{m}}$ to ${\bm{n}}$ \cite{kane_Quantum_2005}, and $\beta_s$ denoting the strength of SOC, causing a band splitting of $2\beta_{so}=6\sqrt{3}\beta_s$.
Besides, the nearest neighbor hopping and the superconducting pairing potential are denoted as $t$ and $\Delta$, respectively.

We assume that the left and right leads are naturally connected, i.e. the coupling only exists in the outmost layer of each lead, and set $t_c(\bm{m},\bm{n})= t$ when $\bm{m}$,$\bm{n}$ are in the outmost layers.
By considering the junction width in $y$ direction to be large,
we can adopt the periodic boundary condition and apply the Bloch theorem along $y$ direction to introduce the transverse wave vector $k_y$.
While $k_y=0$ represents the normal incident case, the nonzero $k_y$ characterizes the process of oblique incidence.
Thus the total Hamiltonian can be rewritten as $H_{tot} = \sum_{k_y} H_{tot}(k_y)$.
By calculating the transport processes of each independent $k_y$ channel and summing them up, we can obtain the whole transport property of this two-dimensional normal metal-Ising superconductor junction.

The spin current flowing from the left lead into the junction can be expressed as \cite{wang_Spinbattery_2004,mao_Charge_2021}
\begin{equation}
    J_s = \frac{\hbar}{2} (J_+ - J_-) ,
\end{equation}
where we consider an arbitrary orientation of spin bias $\hat{\bm{n}}$ and the corresponding basis $\ket{\pm} = \ket{\hat{\bm{n}}\cdot\vec{\bm S}=\pm\hbar/2}$.
In the presence of good quantum number $k_y$, we can derive the total spin current as $J_s=\frac{\hbar}{2}\sum_{k_y}[J_+(k_y)-J_-(k_y)]$.
We apply a spin bias onto the left normal-metal lead and keep the right
Ising superconductor lead under zero bias, so the chemical potentials of the left and right leads are $\mu_{L\pm} =\pm eV$ and $\mu_{R\pm}=0$ with $V$ the spin bias strength.
After choosing the outmost layer of the normal-metal lead to be the center region and expressing the Green's functions $\bm{G}^{r,a}$ and linewidth matrices $\bm{\Gamma}^{L/R}$ in generalized $4\times4$ Nambu basis, we obtain each specific spin current $J_\sigma\ (\sigma=\pm)$ by using nonequilibrium Green's function method \cite{sun_Quantum_2009, lv_Magnetoanisotropic_2018},
\begin{widetext}
\begin{equation}
    J_\sigma = \frac{1}{h} \int \mathrm{d\omega}[
        (f_{L,e\sigma}-f_{L,h\sigma})T_{he,\sigma}
        + (f_{L,e\sigma}-f_{L,e\bar{\sigma}})T_{\bar{e}e,\sigma}
        + (f_{L,e\sigma}-f_R)T_{trans,\sigma}
    ] ,
\end{equation}
\end{widetext}
where $T_{he,\sigma}=\mathrm{Tr}[\bm{\Gamma}^L_{ee\sigma\sigma} \bm{G}^r_{eh\sigma\sigma} \bm{\Gamma}^L_{hh\sigma\sigma} \bm{G}^a_{he\sigma\sigma}]$ is the equal-spin Andreev reflection coefficient representing the process that an incident electron with spin $\sigma$ is reflected as a hole in the same spin sub-band.
$T_{\bar{e}e,\sigma}=\mathrm{Tr}[\bm{\Gamma}^L_{ee\sigma\sigma} \bm{G}^r_{ee\sigma\bar{\sigma}} \bm{\Gamma}^L_{ee\bar{\sigma}\bar{\sigma}} \bm{G}^a_{ee\bar{\sigma}\sigma}]$ is the spin-flip reflection coefficient, with $\bar{\sigma}$ indicating the opposite spin of $\sigma$.
$T_{trans,\sigma}=\mathrm{Tr}[\bm{\Gamma}^L_{ee\sigma\sigma}{(\bm{G}^r\bm{\Gamma}^R\bm{G}^a)}_{ee\sigma\sigma}]$ is the transmission coefficient corresponding to the quasiparticle tunneling.
The spin bias setting requires $f_{L,e\sigma}=f_{L,h\bar{\sigma}}=f(\omega-eV)$ and $f_{L,e\bar{\sigma}}=f_{L,h\sigma}=f(\omega+eV)$ in the normal-metal lead and $f_R = f(\omega)$ in the Ising superconductor lead, with $f(\omega)$ being the Fermi distribution function.
Besides, as the sign of the spin $\sigma$ only affects the phases of the amplitudes of the equal-spin Andreev reflection and spin-flip reflection, it would not change the coefficients.
Therefore, we simply omit the subscript $\sigma$ of the $T_{he}, T_{\bar{e}e}$ in the discussions below.

\section{\label{equal-spin}Equal-spin Andreev reflection}
As demonstrated in previous works \cite{zhou_Ising_2016,lv_Magnetoanisotropic_2018,tang_Magnetic_2021},
the pairing symmetry in Ising superconductors can be expressed as pairing correlation $\bm{F}=\Delta[\psi_s\sigma_0 + \bm{d}\cdot\bm{\sigma}]i\sigma_y$, with a scalar $\psi_s$ parametrizing the spin-singlet pairing correlation and a vector $\bm{d}$ parametrizing the spin-triplet pairing.
Near the $\mathbf{K}$ valley, $\bm{d}=(0,0,d_z)$ with
\begin{eqnarray}\label{eq_dz}
    &&d_{z}(\bm{p}+\mathbf{K},\omega)
    = \nonumber\\
    &&\frac{2\beta_{so}\xi_{\bm{p}}}
    {[\omega^2_+ - {(\xi_{\bm{p}} + \beta_{so})}^2 - \Delta^2][\omega^2_+ - {(\xi_{\bm{p}} - \beta_{so})}^2 - \Delta^2]},
\end{eqnarray}
where $\beta_{so} = 3\sqrt{3}\beta_s, \omega_+=\omega+i0^+$, and $\xi_{\bm{p}}=|\bm{p}|^2/2m-E_F$ is the kinetic energy measured from the Fermi surface with $m$ the effective mass \cite{lv_Magnetoanisotropic_2018}.
We thus find that $d_z$ vanishes without SOC ($\beta_s$).
By choosing an arbitrary orientation of the spin quantization axis $\hat{\bm n}=(\sin\theta\cos\phi,\sin\theta\sin\phi,\cos\theta)$ with the azimuth angles $\theta$ and $\phi$ [see Fig. \ref{p_model}(a)],
we further get the equal-spin-triplet pairing correlation as $F_{\sigma\sigma}=\sigma d_z\sin\theta e^{i\phi} ,\ \sigma=\pm$.
It shows that the spin of Cooper pair has finite component in $xOy$ plane but none in the $z$ direction.
Such Cooper pairs carry both charge and spin,
so they can transport not only charge supercurrent but also spin supercurrent \cite{mao_Spin_2022}.

\begin{figure}[htbp]
    \includegraphics[width=\linewidth]{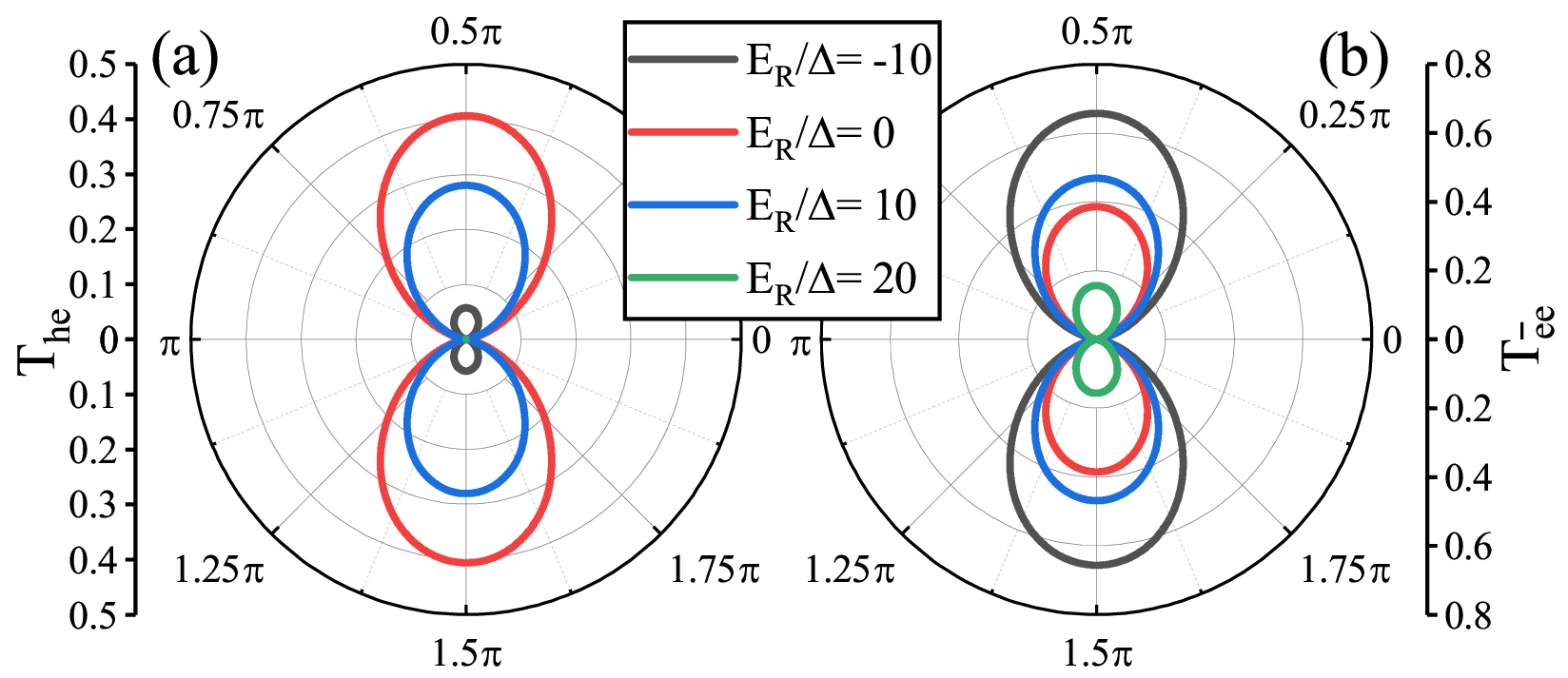}
    \caption{\label{p_angle}
    Polar angle $\theta$ dependence of equal-spin Andreev reflection coefficient $T_{he}$ (a) and spin-flip reflection coefficient $T_{\bar{e}e}$ (b) with different on-site energy $E_R$.
    We keep $\Delta=1\mathrm{meV},\beta=\epsilon=-1\mathrm{eV},t=-3\mathrm{eV},\beta_s=2\mathrm{meV},E_L=20\mathrm{meV},k_y=0,\omega=0$ and $\phi=0$ in calculations.}
\end{figure}

The nonzero $F_{\sigma\sigma}$ indicates that it is possible to generate and inject spin-triplet Cooper pairs into Ising superconductors to achieve dissipationless spin transport.
This is done through the equal-spin Andreev reflection process \cite{visani_Equalspin_2012,mao_Charge_2021,lv_Magnetoanisotropic_2018} 
[see Fig. \ref{p_model}(b)].
The strength of $T_{he}$ determines the magnitude of the spin supercurrent and is related to the equal-spin-triplet pairing correlation $F_{\sigma\sigma}$, showing a strong anisotropic angular dependence.
Fig. \ref{p_angle}(a) shows how $T_{he}$ depends on the spin direction of incident electrons.
When the spin bias is along $z$ direction, $F_{\sigma\sigma}|_{\theta=0}=0$.
The zero equal-spin-triplet pairing correlation indicates that all the Cooper pairs are without z-direction spin component.
Thus, none can carry z-direction spin supercurrent and $T_{he}$ disappears.
As the orientation deviates from $z$ axis, $T_{he}$ rises up, which is consistent with the $F_{\sigma\sigma}-\theta$ relation, indicating the increase of spin-triplet Cooper pairs.
Both $T_{he}$ and $F_{\sigma\sigma}$ reach their maxima when the spin orientation is in $xOy$ plane, where only Cooper pairs formed by equal-spin electrons contribute to the spin-triplet pairing correlation.
Due to the dissipationless nature of Cooper pairs, it is beneficial to long-distance spin transport.
Moreover, as the azimuth angle $\phi$ only attaches a phase factor to $F_{\sigma\sigma}$, it does not affect $T_{he}$ and the transport process.
\begin{figure}[htbp]
    \includegraphics[width=\linewidth]{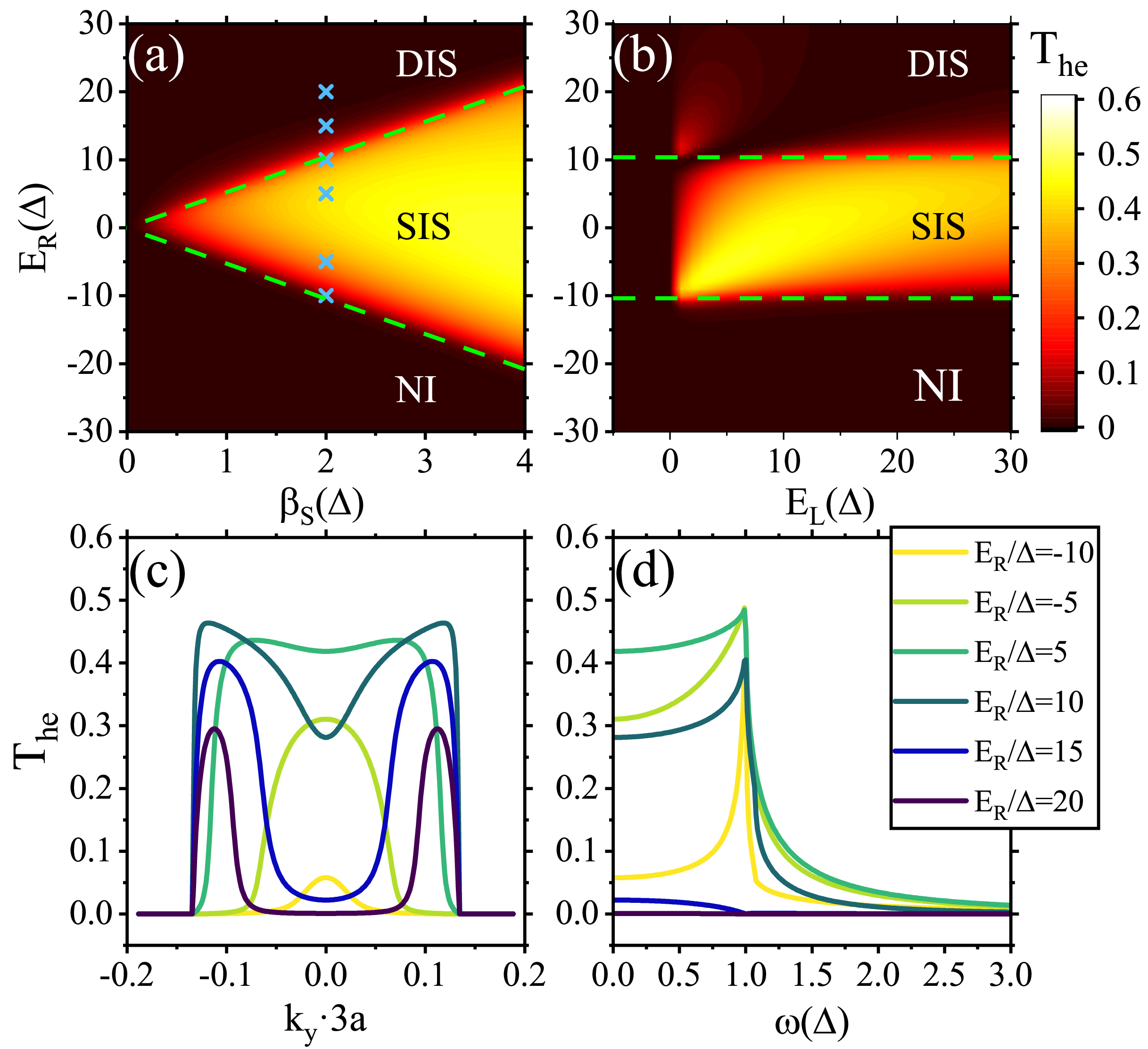}
    \caption{\label{p_T1}
    (a) and (b) are the color-coded equal-spin Andreev reflection coefficient $T_{he}$ as functions of SOC strength $\beta_s$ and on-site energy $E_{L/R}$.
    The phases separated by dashed lines are normal insulating (NI) state , single-band Ising superconducting (SIS) state and double-band Ising superconducting (DIS) state, respectively.
    (c) and (d) are the equal-spin Andreev reflection coefficient $T_{he}$ as functions of transverse wave vector $k_y$ and incident energy $\omega$ with different on-site energy $E_R$.
    The starting points in (c) ($k_y=0$) and (d) ($\omega=0$) correspond to the cross dots in (a).
    We keep $\theta=\pi/2$ in calculations while other parameters are the same as those in Fig. \ref{p_angle}.}
\end{figure}

Besides, the equal-spin Andreev reflection process also depends on the SOC strength $\beta_s$ and the on-site energy $E_{R/L}$ [shown in Fig. \ref{p_T1}(a) and \ref{p_T1}(b)], which splits the energy band and affects the Fermi surface in the Ising superconductor, respectively.
As we consider the normal incident case ($k_y=0$) with zero incident energy, for a fixed $\beta_s$, the increasing of $E_R$ shifts the Fermi surface below the bottom of bands to the middle of two spin sub-bands or even higher [see Fig. \ref{p_model}(a)], making the right lead experience a phase transition from normal insulating (NI) state through single-band Ising superconducting (SIS) state to double-band Ising superconducting (DIS) state.
The transition lines are given as
\begin{eqnarray}\label{eq_transline1}
    E_R = \pm  3\sqrt{3} \beta_{s}.
\end{eqnarray}
In different phases, the strength of the equal-spin Andreev reflection $T_{he}$ appears to be quite different.
When the right lead is in the NI state, no carrier can flow through it so that $T_{he}$ disappears 
[see Fig. \ref{p_T1}(a) and \ref{p_T1}(b)].
As $E_R$ increases, the magnitude of $T_{he}$ is first dramatically enhanced in the SIS phase but then suppressed in the DIS phase 
[Fig. \ref{p_T1}(a) and \ref{p_T1}(b)].
These behaviors can be qualitatively understood from $|d_z|$ that has two extrema at $\xi_{\bm{p}} \approx \pm\beta_{so}$.
If we consider the normal incident case ($k_y=0$), these extrema appear at
\begin{eqnarray}\label{eq_extre1}
    E_R =  \frac{k_x^2}{2m}\pm3\sqrt{3}\beta_{s}.
 \end{eqnarray}
For the transport processes, $k_x$ varies, passing both extrema if $E_R > 3\sqrt{3}\beta_{s}$, one extremum if $-3\sqrt{3}\beta_{s}\leq E_R\leq 3\sqrt{3}\beta_{s}$ or no extremum if $E_R < -3\sqrt{3}\beta_{s}$.
Only the second case can introduce large $d_z$ and ensure the arising of remarkable equal-spin Andreev reflection process in the SIS phase.
Meanwhile, for a large $\xi_{\bm{p}}$, $d_z$ undergoes a cubic decay along with the increase of $\xi_{\bm{p}}$ 
[see Eq. (\ref{eq_dz})].
This means that few spin-triplet Cooper pairs exist apart from the extrema, resulting in $T_{he}$ being small in both the DIS and NI phases.

Further calculations tell us more characteristics of the equal-spin Andreev reflection $T_{he}$
that it can be influenced by the Fermi wavelength mismatch in two leads, the transverse wave vector $k_y$ and the incident energy $\omega$.
As shown in Fig. \ref{p_T1}(b), when $E_L<0$, the Fermi surface in the left lead locates in the energy gap, making it an insulating state.
Therefore, no electron can transport through the left lead, causing the zero value of $T_{he}$.
As $E_L$ varies greater than zero, it affects the Fermi wavelength $k_{F,L}$ in the left lead so that $T_{he}$ can get enhancement at some appropriate $E_L$.
In other cases, $T_{he}$ suffers a decrease due to the Fermi wavelength mismatch in two sides \cite{lv_Magnetoanisotropic_2018}.

We then select cross dots from Fig. \ref{p_T1}(a) as the starting points to investigate the influence of oblique incidence and nonzero incident energy.
The results are plotted in the Fig. \ref{p_T1}(c) and \ref{p_T1}(d).
For oblique incident cases, the incoming electrons have nonzero transverse momentum, corresponding to nonzero $k_y$.
At some large $|k_y|>|k_{F,L}|$, $T_{he}$ vanishes because there exists no electron in the left lead.
Otherwise, we can rewrite the extrema condition as
\begin{eqnarray}\label{eq_extre2}
    \ E_R-\frac{k_y^2}{2m} = \frac{k_x^2}{2m}\pm 3\sqrt{3}\beta_{s}.
 \end{eqnarray}
Under the same argument with Eq. (\ref{eq_extre1}),
we can get notable $T_{he}$ values at region $-3\sqrt{3}\beta_{s}\leq E_R-k_y^2/2m\leq 3\sqrt{3}\beta_{s}$.
It means that the oblique equal-spin Andreev reflection process can occur even in the DIS phase, consistent with the numerical results in Fig. \ref{p_T1}(c).
For varying incident energy $\omega$, Fig. \ref{p_T1}(d) shows that $T_{he}$ decays quickly outside the superconducting gap $\Delta$, regardless of the value of $E_R$.
Inside the superconducting gap $\Delta$, the increase of $\omega$ can effectively enhance the strength of $T_{he}$ in the SIS phase but suppress the process in the DIS phase.
This enhancement can also be understood from $d_z$ as its absolute value increases with $\omega$ approaching $\Delta$ inside the gap.

\section{\label{spin-flip}Spin-flip reflection}
\begin{figure}[htbp]
    \includegraphics[width=\linewidth]{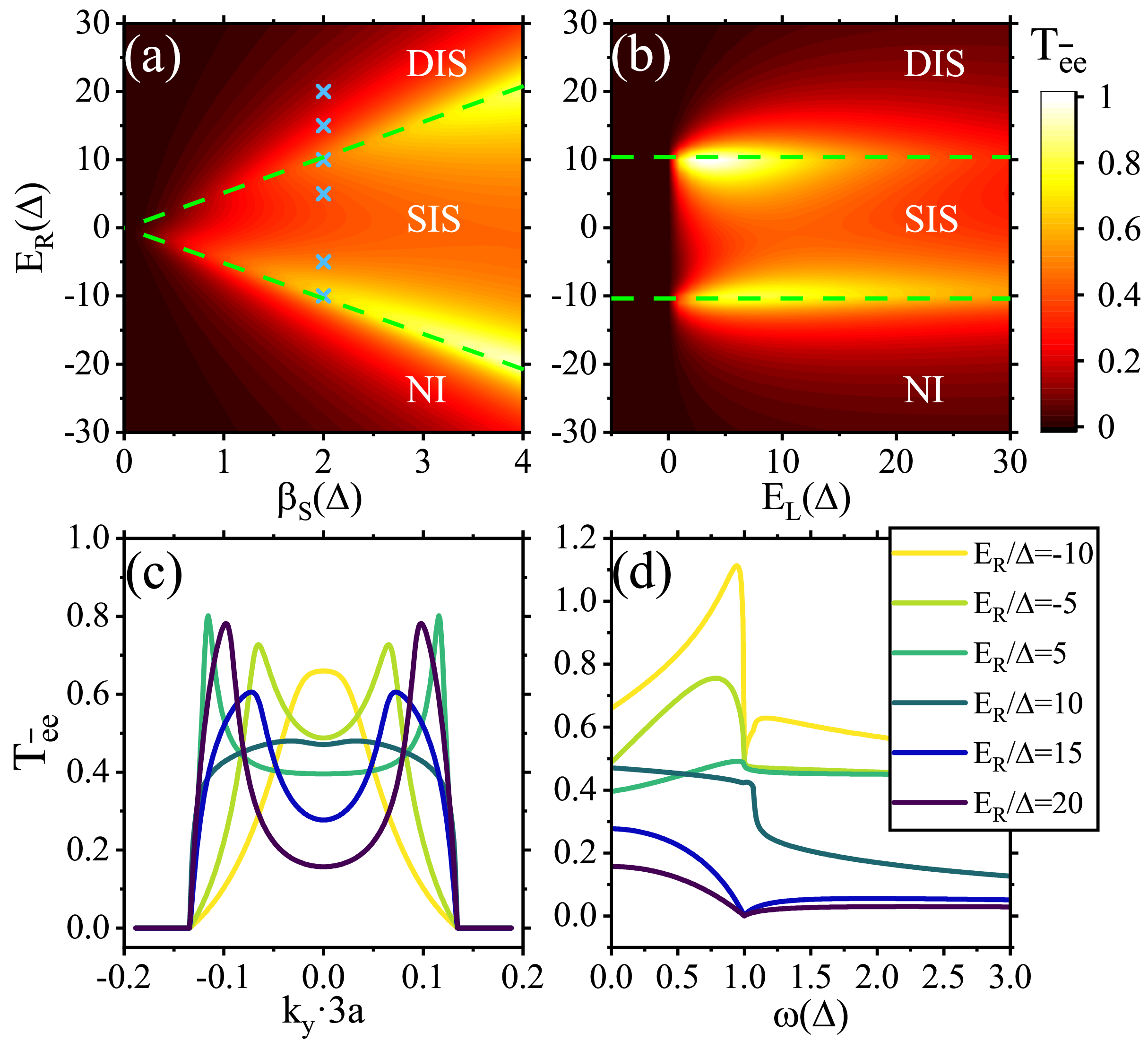}
    \caption{\label{p_T2}
    (a) and (b) are the color-coded spin-flip reflection coefficient $T_{\bar{e}e}$ as functions of SOC strength $\beta_s$ and on-site energy $E_{L/R}$.
    (c) and (d) are the spin-flip reflection coefficient $T_{\bar{e}e}$ as functions of transverse wave vector $k_y$ and incident energy $\omega$ with different on-site energy $E_R$.
    The starting points in (c) ($k_y=0$) and (d) ($\omega=0$) correspond to the cross dots in (a).
    We keep $\theta=\pi/2$ in calculations while other parameters are the same as those in Fig. \ref{p_angle}.}
\end{figure}

Next we turn to another process associated with spin transport, that is the spin-flip reflection.
Although it always contributes to the spin current in the normal-metal side, whether this current can flow into the other side depends on carriers.
In exciton spin superconductors \cite{sun_Spin_2011,liu_Spontaneous_2012,sun_Spinpolarized_2013}, the carrier is electron-hole pair that form a dissipationless superfluid state, which makes the spin-flip reflection act as an Andreev reflection.
Therefore, in the metal-spin superconductor interface, such reflection injects spin supercurrent into the spin superconductor by inducing electron-hole pairs \cite{wang_Spin_2005,lv_Spinflip_2017}.
However, in Ising superconductors, those who carry the spin supercurrent are spin-triplet Cooper pairs generated only through the equal-spin Andreev reflection.
Thus the spin-flip reflection process cannot inject the spin into the Ising superconductor but dissipates it around the interface of the junction.
Calculated results about spin-flip reflection coefficient $T_{\bar{e}e}$ are shown in Fig. \ref{p_angle}(b) and Fig. \ref{p_T2}.
$T_{\bar{e}e}=0$ when the polar angle $\theta=0, \pi$
and it reaches the maximum at $\theta=\pi/2, 3\pi/2$, which is
similar to the equal-spin Andreev reflection $T_{he}$.
But the behavior of $T_{\bar{e}e}$ as functions of $\beta_S$ and $E_R$ ($E_L$ and $E_R$) 
[see Figs. \ref{p_T2}(a) and \ref{p_T2}(b)] 
is quite different from $T_{he}$ 
[see Figs. \ref{p_T1}(a) and \ref{p_T1}(b)].

The spin-flip reflection can be understood from the torque provided by the SOC-induced effective Zeeman field.
As the effective field is along $z$ direction, the incident electrons can feel the torque as soon as their spin orientations are apart from $z$ axis.
Such effect reaches its maximum when the spin orientation is perpendicular to $z$ axis, namely along the $xOy$ plane, which explains the anisotropic angular dependence 
[see Fig. \ref{p_angle}(b)].
Moreover, due to the existence of SOC, the spin-flip process can happen even in the NI phase and shows two maximum peaks around the transition lines 
[see Fig. \ref{p_T2}(a) and \ref{p_T2}(b)].
While $T_{he}$ appears to be larger near the NI-SIS line, the peak of $T_{\bar{e}e}$ near the NI-SIS line is however lower than that near the SIS-DIS line, showing the different influence of $E_L$ on two processes.

Our calculations further reveal that for oblique incident cases, the spin-flip reflection in both SIS and DIS phases are enhanced at some high $|k_y|$ 
[see Fig. \ref{p_T2} (c)].
This is because the transition lines satisfy
\begin{eqnarray}\label{eq_transline2}
    E_R-\frac{k_y^2}{2m} = \pm3\sqrt{3}\beta_{s}
\end{eqnarray}
for a nonzero $k_y$.
As $E_R$ varies, the peaks correspond to a higher $|k_y|$.
Besides, for different incident energy $\omega$, the behavior of the spin-flip reflection $T_{\bar{e}e}$ differs slightly from that of
the equal-spin Andreev reflection $T_{he}$ 
[see Fig. \ref{p_T2} (d)].
Inside the superconducting gap, the suppression of $T_{\bar{e}e}$ happens not only in the DIS phase but also near the SIS-DIS transition region.
As $\omega$ varies greater than $\Delta$, unlike $T_{he}$ decaying quickly to zero, $T_{\bar{e}e}$ remains a finite strength and is only suppressed in or near the DIS phase, showing that the spin-flip reflection is contributed by SOC rather than superconducting.

\section{\label{spin-cond}Spin conductance}
We then investigate the total spin current of the system, considering the differential spin conductance in the small spin bias limit and finite spin bias cases.
\begin{figure}[htbp]
    \includegraphics[width=\linewidth]{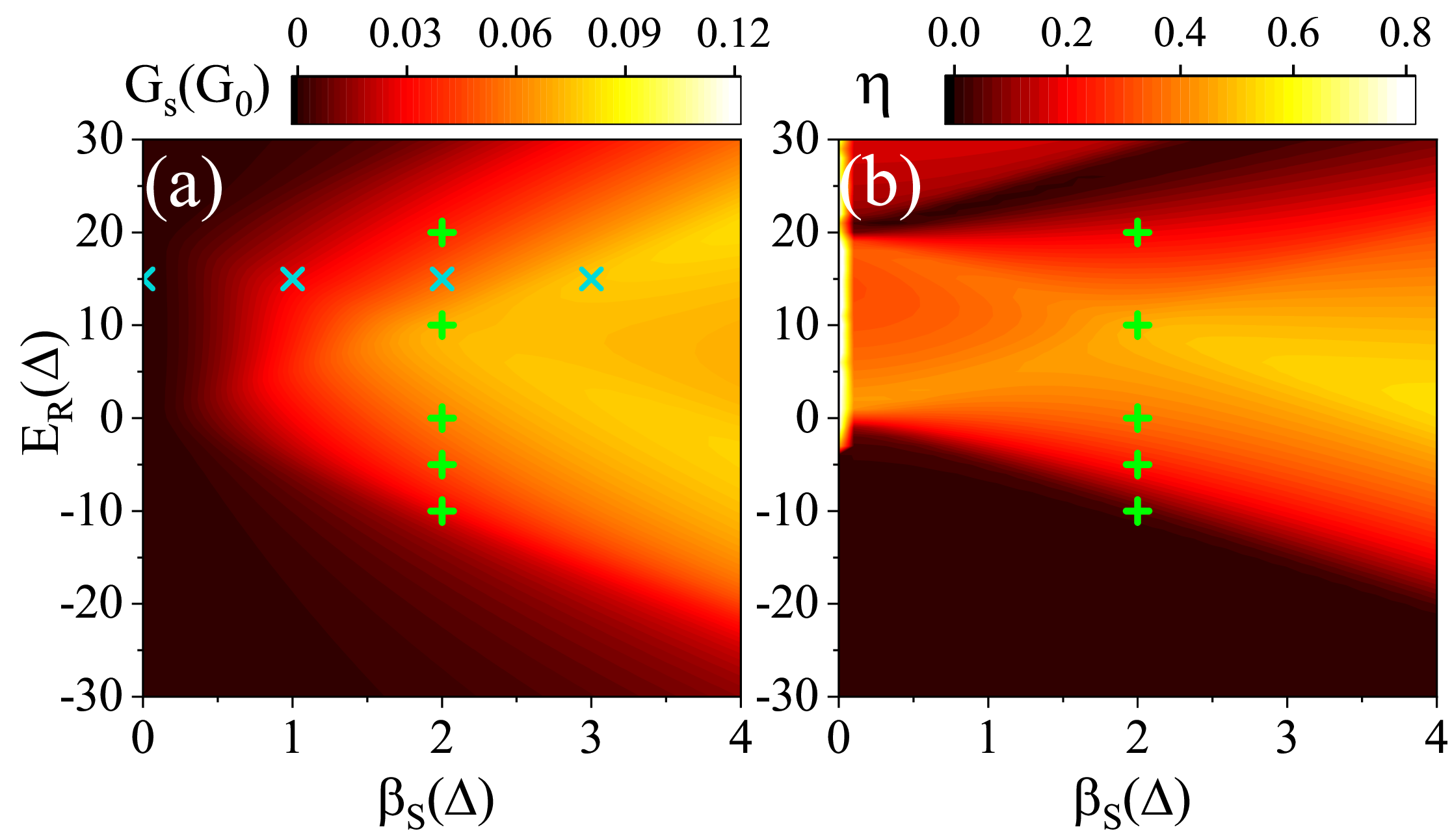}
    \caption{\label{p_G}
    (a) is the calculated results of linear spin conductance $G_s$ as 
    a function of $E_R$ and $\beta_s$. 
    Here $G_0=e/2\pi\cdot W/3a$ with $a$ the 
    nearest-neighbor site-site distance and $W$ the ribbon width.  
    We select vertical and transverse dots to further investigate the characteristics of $G_s$ 
    [shown in Fig. \ref{p_finite_G} (a) and \ref{p_finite_G} (b) respectively].
    (b) is the spin injection efficiency $\eta$ as a function of $E_R$ and $\beta_s$.
    We select vertical dots to further investigate the characteristics of $\eta$ [shown in Fig. \ref{p_finite_G} (d)].
    We keep $\theta=\pi/2$ and sum up all the $k_y$ channels in calculations while other parameters are the same as those in Fig. \ref{p_angle}.}
\end{figure}

The differential spin conductance is defined as $G_s=\mathrm{d}J_s/\mathrm{d}V$.
In small spin bias limit where $V\rightarrow 0$, it can be represented by $G_s =\sum_{k_y}(e/2\pi)[T_{he,+}+T_{he,-}+T_{\bar{e}e,+}+T_{\bar{e}e,-}]$ for the absence of the quasiparticle tunneling process.
This result presents the contribution of equal-spin Andreev reflection and spin-flip reflection to the spin conductance.
As shown in Fig. \ref{p_model}(b), considering the spin bias in $x$ direction as an example, the occupied spin $x$ channel injects Cooper pairs with spin $S=1, S_x=1$ into the right lead, while those Cooper pairs carrying $S=1, S_x=-1$ flow out of the Ising superconductor to the unoccupied spin $\bar{x}$ channel of the left lead, both with the help of the equal-spin Andreev reflection.
Therefore, Cooper pairs carrying opposite spin flow oppositely in the Ising superconductor, causing a dissipationless spin supercurrent.
Meanwhile, the spin-flip process can dissipate the spin current near the interface of the junction, consuming the injected spin current from the left side and offering a channel to contribute to the spin conductance.
Next we further consider the quasiparticle tunneling process occurring in the finite bias.
Compared with the enhancement of spin lifetime of quasiparticles in normal superconductors \cite{linder_Superconducting_2015}, the situation here is more complicated.
On one hand, the spin-orbit scattering would flip the spin of quasiparticles, on the other hand, the spin-triplet correlations induced by SOC gives the way that the recombination of quasiparticles would also induce spin-triplet Cooper pairs and carry a spin supercurrent \cite{seja_Quasiclassical_2021}.
Our calculated results show in Fig. \ref{p_G}(a) that although the spin transport for the normal incidence ($k_y=0$) mainly
occurs in the SIS phase, the linear spin conductance $G_s$ is still appreciable in the DIS phase for the enhancement contributed by oblique incidence.

As discussed above, this spin conductance is defined in the left lead and is usually unequal to the spin supercurrent in the Ising superconductor.
Therefore, to investigate the spin supercurrent flowing in the Ising supercondcutor, we further define the spin injection efficiency as $\eta=1-G_{s,dis}/G_{s,tot}$, where $G_{s,dis}$ is the dissipated part, which is governed by the spin-flip reflection, and $G_{s,tot}$ is the total spin current in the normal-metal lead.
One can see from Fig. \ref{p_G}(b) that in the DIS phase, $\eta$ is a relatively small value and most of the spin current flowing through normal-metal lead will be dissipated near the interface.
This is consistent with the result in Fig. \ref{p_T1} that the equal-spin Andreev reflection is suppressed in the DIS phase.

\begin{figure}[htbp]
    \includegraphics[width=\linewidth]{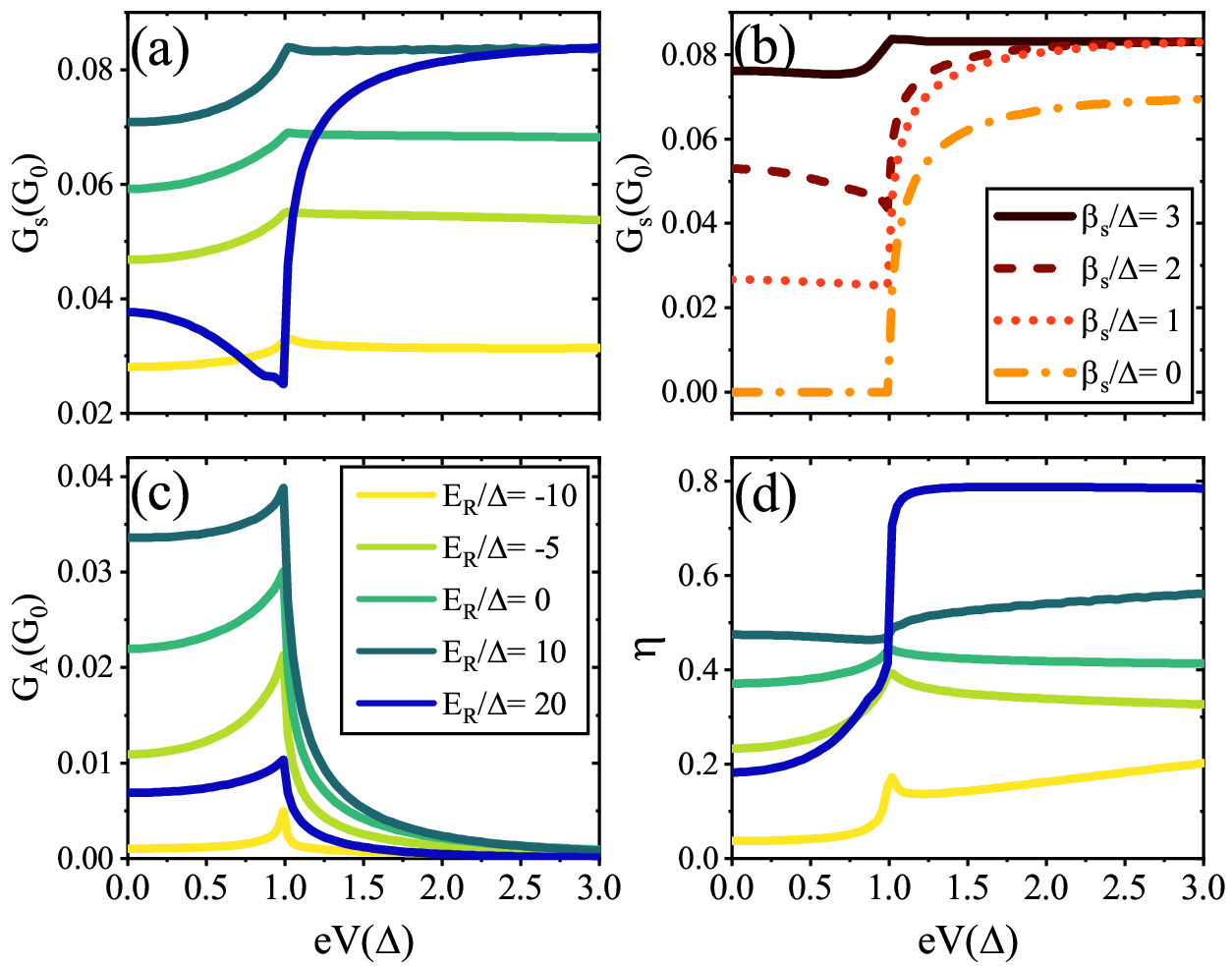}
    \caption{\label{p_finite_G} Results of differential spin conductance $G_s$ in finite spin bias cases with (a) fixed $\beta_s=2\mathrm{meV}$ and (b) fixed $E_R=15\mathrm{meV}$.
    (c) is the differential spin conductance $G_A$ contributed by equal-spin Andreev reflection and (d) is the spin injection efficiency $\eta$, both fixing $\beta_s=2\mathrm{meV}$ in calculations.
    While (a) (c) (d) use the same legend in (c), we keep $\theta=\pi/2$ in calculations and other parameters are the same as those in Fig. \ref{p_angle}.}
\end{figure}

Starting from $G_s$ in small spin bias limit, we further calculate the spin conductance under finite spin bias
[see Fig. \ref{p_finite_G}(a) and \ref{p_finite_G}(b)].
When the spin bias varies from zero to $\Delta$, 
the spin conductance $G_s$ increases slightly in the SIS phase, 
but decreases in the DIS phase 
[see the curve with $E_R/\Delta =20$ in Fig. \ref{p_finite_G}(a)].
Out of the superconducting gap, $G_s$ decreases slightly in the SIS phase
and increases in the DIS phase with the increase of spin bias $V$.
When $V>2\Delta$, $G_s$ remains nearly unchanged.
If we further look at the part contributed by the equal-spin Andreev reflection 
[denoted as $G_A$ in Fig. \ref{p_finite_G}(c)], 
we find that it always grows up inside the superconducting gap but decays quickly at $eV>\Delta$.
This is similar to the normal Andreev reflection coefficient in normal lead-superconductor heterostructures, where a resonant Andreev reflection happens near the gap edges \cite{blonder_Transition_1982}.
Meanwhile, the calculated results of spin injection efficiency $\eta$ in Fig. \ref{p_finite_G}(d) show that no matter $eV$ is smaller or larger than $\Delta$, in the SIS phase, $\eta$ is almost unchanged, but in the DIS phase, $\eta$ grows quickly as the spin bias approches the superconducting gap and remains a relatively high level at $eV>\Delta$.
This large $\eta$ is due to the large double-band-induced transmission coefficients $T_{trans,\sigma}$, together with the limited interface dissipation in the DIS phase.
Similar to the current-to-superflow conversion in charge transport \cite{seja_Quasiclassical_2021}, we expect here a conversion from spin current carried by quasiparticles to spin supercurrent carried by spin-triplet Cooper pairs. 
So the spin injection by $T_{trans,\sigma}$ can convert into the dissipationless spin supercurrent. 
On the other hand, when $eV<\Delta$, the spin injection efficiency $\eta$ has a larger value in the SIS phase than in the NI and DIS phases. 
Here $\eta$ in the SIS phase can exceeds 40\% at the suitable parameter.
In particular, for $eV<\Delta$, the spin injection efficiency is 
completely contributed by the equal-spin Andreev reflection 
and the spin current is carried by the spin-triplet Cooper pairs, 
so it is undoubtedly non-dissipative.

\section{\label{diss}Discussion and conclusions}
The calculations above show the transport characteristics of spin superconductivity in Ising superconductors, as well as the possibility of detecting spin superconducting state.
As both spin supercurrent and spin dissipation can be controlled by various parameters, it is convenient in experiments to adjust the transport by electrical gate controlling.
Moreover, as the spin-triplet component can be regarded as the combination of ferromagnetism and superconducting order parameter,
in traditional superconducting spintronics,
this is usually obtained by the proximity between a normal superconductor and a ferromagnet \cite{linder_Superconducting_2015,keizer_spin_2006,gomperud_Spin_2015,montiel_Generation_2018,jeon_Enhanced_2018,eschrig_Spinpolarized_2015}.
However, the spin-triplet Cooper pairs here in Ising superconductors are intrinsic quantum coherent states, providing unique advantages in long-distance spin transport.
As the spin transport highly depends on spin directions and gate voltages, it makes Ising superconductors competitive in the aspect of controlled long-distance spin transport.

In summary, we show that Ising superconductors also have the characteristic of spin superconductivity, as its spin-triplet component 
can carry spin supercurrent.
Using the nonequilibrium Green's function method, we comprehensively investigate the spin transport in the normal metal-Ising superconductor junction.
The calculations show that spin supercurrent can be injected into the Ising supercondcutor by equal-spin Andreev reflection process, while some spin current is dissipated near the interface of the junction by spin-flip process.
We emphasize that the characteristic of spin superconductivity in Ising superconductors makes it useful to controlled long-distance dissipationless spin transport, which can promote the further development of spintronics.

\begin{acknowledgments}

This work was financially supported by NSF-China (Grant No. 11921005), 
National Key R and D Program of China (Grant No. 2017YFA0303301), and the Strategic priority Research Program of Chinese Academy of Sciences (Grant No. XDB28000000).

\end{acknowledgments}

\bibliography{ref1}

\end{document}